\documentclass[10pt,twocolumn,letterpaper]{article}
\usepackage[margin=2.5cm]{geometry}
\usepackage{times,epsfig,graphicx,amsmath,amssymb}
\usepackage[breaklinks=true,colorlinks=true,bookmarks=false]{hyperref}
\date{}
\usepackage{caption}
\usepackage{adjustbox}

\title{\textbf{Densely Connected Recurrent Residual (Dense R2UNet) Convolutional Neural Network for Segmentation of Lung CT Images}}

\author{%
Kaushik Dutta\\
Department of Radiology, Washington University School of Medicine \\
{\tt kaushik.dutta@wustl.edu}
}

\begin{document}
\maketitle

\begin{center}\textbf{Abstract}\\~\\\parbox{0.475\textwidth}{\em
    
   Deep Learning networks have established themselves as providing state of art performance for semantic segmentation. These techniques are widely applied specifically to medical detection, segmentation and classification. The advent of the U-Net based architecture has become particularly popular for this application. In this paper we present the Dense Recurrent Residual Convolutional Neural Network(Dense R2U CNN) which is a synthesis of Recurrent CNN, Residual Network and Dense Convolutional Network based on the U-Net model architecture. The residual unit helps training deeper network, while the dense recurrent layers enhances feature propagation needed for segmentation. The proposed model tested on the benchmark Lung Lesion dataset showed better performance on segmentation tasks than its equivalent models. 

}\end{center}

\section{Introduction}
In recent times the convolutional neural network have become the state of art machine learning approach for image classification\cite{NIPS2012_c399862d}, segmentation\cite{long2015fully} and detection\cite{wang2015transferring}. Although introduced around two decades ago\cite{lecun1989backpropagation} deep learning has become increasingly popular application in recent times due to modern hardware like GPUs which provides the computation power required for the training of multilayered CNNs. Moreover the deep neural networks learns high level features from data thus eliminating hard feature extraction as compared to traditional machine learning. The neural networks utilizes efficient optimization techniques that helps building a very deep and stable network. Several popular deep CNNs have been proposed in the past decade like AlexNet\cite{NIPS2012_c399862d}, VGG\cite{simonyan2014very}, InceptionNet\cite{szegedy2015going}, DenseNet\cite{huang2017densely}, ResNet\cite{he2016deep} etc which are used for semantic segmentation. One of the biggest challenge for training deep networks is that it requires massive amounts of labeled data as ground truth for training but recently many datasets like ImageNet\cite{deng2009imagenet} has become publicly available for training of networks. 
\newline
Motivated by the great success of deep learning approaches in solving computer vision problems, different approaches of it is applied in medical imaging for segmentation, classification, disease detection and registration. Medical image segmentation is an important area in diagnosis, monitoring and treatment of diseases. In medical imaging manual segmentation by experts across different imaging modalities is still considered as a gold standard for diagnosis and treatment. Manual segmentation is tedious slow and prone to human error. Deep learning based techniques are applied in cell segmentation, segmentation of different parts of organs like brain\cite{menze2014multimodal}, lung\cite{kalinovsky2016lung}, pancreas\cite{roth2015deeporgan}, prostate\cite{yu2017volumetric} and also for tumor lesion segmentation\cite{havaei2017brain} in various organs. However, there are few challenges in medical image segmentation like scarcity of ground truth and class imbalance. To overcome the problem of scarcity of annotated data for training, data augmentation and transformation is applied to increase labeled samples. In addition patch based approaches are utilized to counter the class imbalance problem.
\newline
The CNN model architecture used for segmentation tasks requires both convolutional encoding and decoding units as depicted in U-Net\cite{ronneberger2015u}. The encoding unit of the network is used to encode the input images into a large number of unique feature maps of lower dimensionality(which decreases down the network as feature maps increases). The decoding unit performs deconvolution i.e produces segmentation probability maps from the encoded feature maps. It also utilized skip connections which can directly transfer information from the encoding portion of the network to the decoding portion of the network. One of the major goals of segmentation networks are to increase the performance of the network without increasing the network parameters.
\newline
In this paper we propose a modified architecture based on the native U-Net model utilizing recurrent residual convolutional networks with dense interconnections to connect all layers and ensure maximum information flow between the layers. The network is trained and tested for Lung CT segmentation tasks in an end-to-end image based approach and segmentation accuracy is compared to recent state-of-art methods.

\section{Background \& Related Work}
The designing of optimized network architecture has been an integral focus of neural network research since its inception. Advanced deep learning networks nowadays have hundreds of layers and millions of parameters which requires huge computation. Thus there is a need to explore the most optimized architecture and connectivity patterns in a network. Semantic segmentation is an active research area where CNNs are used for pixel classification and DeepLab\cite{chen2017deeplab} is one such state of art algorithm. The Seg-Net\cite{badrinarayanan2017segnet} consists of an encoding network of that of a 13 layer VGG16\cite{simonyan2014very} and a decoding layer for pixel wise classification. The most popular network that is used for biomedical image segmentation is the 'U-Net'\cite{ronneberger2015u} architecture which has a symmetrical convolutional encoding and decoding(down-convolution) units. The convolution operation is followed by ReLU\cite{szegedy2015going} activation layer in both units. U-Net had a few advantages than other network, it provides both location and contextual information, works fine with limited datasets and preserves the full context of the input images as compared to patch based segmentation. The UNet also uses skip connections\cite{drozdzal2016importance} where the features are directly taken from the encoding layers and concatenated to the decoding layer. 
\newline
There has been many variants of the U-Net like the ResUNet, DenseUNet, VNet and the R2Unet which all uses the basic U-Net architecture with different connectivity patterns. The ResUnet\cite{he2016deep} utilizes deep residual learning framework which eases the training of deeper network by tackling the vanishing gradient problems. In ResUnet the results from the convolutional layers are combined with the input of the layer to constitute the input to the next layer. High-Res3DNet\cite{li2017compactness} is used for 3D segmentation tasks using residual connections. The DenseUnet\cite{kolavrik2019optimized} was developed for high resolution 3D segmentation of brain and spine by implementing densely connected layers between the convolution operations. The V-Net\cite{milletari2016v} network is used for volumetric segmentation with fully connected layers and residual networks. There is also a Cascaded V-Net\cite{hao2019two} which implements multiple V-Nets cascaded in series for improved performance. In 2016 VoxResNet was proposed which does a voxel wise segmentation of the brain by using residual networks and feature map summation across different layers. Alom et al. proposed two variant of UNet i.e Recurrent convolutional neural network(RCNN) and Recurrent Residual CNN (R2CNN) both of which uses operations of the recurrent convolutional layers at discrete timesteps expressed according to the RCNN. In our work we have used the recurrent convolutional layers along with residual layers and dense interconnections and named it Dense-R2UNet.

\begin{figure*}[!t]
\begin{center}
\includegraphics[height=30em]{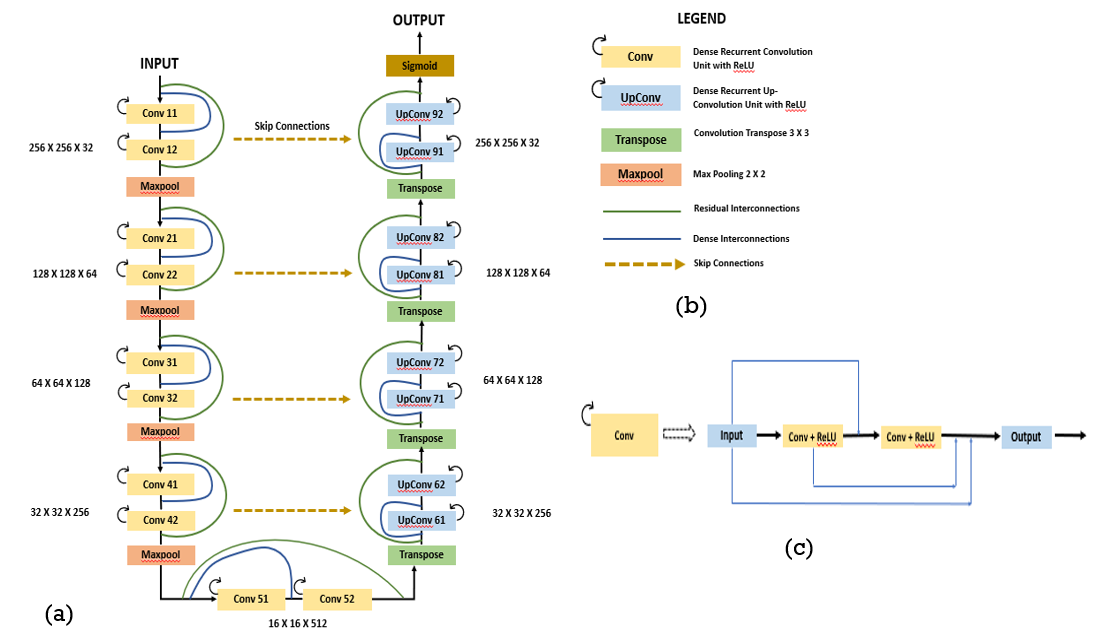}
\end{center}
   \caption{(a) Network Architecture of Dense R2UNet (b) Legend signifying the different units of the network (c) Dense Recurrent Residual Convolutional Block with t=2}
\label{fig:onecol}
\end{figure*}
\begin{figure}[!t]
	\begin{center}
		\includegraphics[width = 23.5em,height=9.5em]{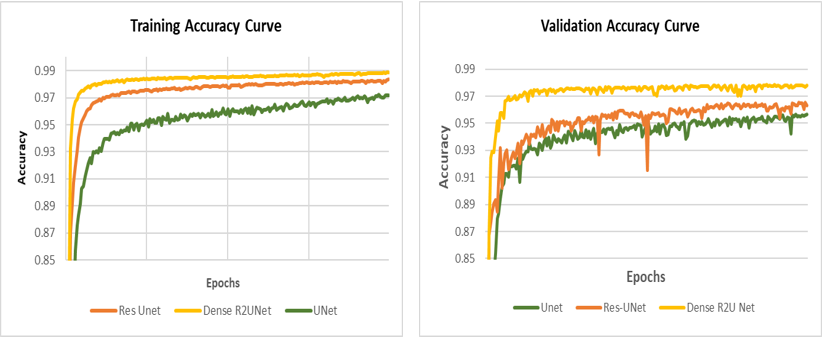} 
	\end{center}
	\caption{Training Accuracy Curve and Validation Accuracy Curve for Lung Segmentation using UNet, ResUNet and Dense R2UNet}
	\label{fig:twocol}
\end{figure}

\section{Proposed Approach}
\subsection{Dataset - Lung Segmentation}
We used the Lung Nodule Analysis(LUNA) dataset which was used in the competition at the Kaggle Data Science Bowl 2017 to find the lung lesions in 2D and 3D CT images. The online dataset consists of 534 2D samples along with their respective labels for lung segmentation\cite{WinNT}. For our experiment we used 80\% of the samples for training and validation and 20\% of the dataset was put aside for testing. The 2D samples were resized from their original size of 512 X 512 to 256 X 256 for implementation in our network. 
\subsection{Network Architecture}
The architecture of Dense R2UNet is based on three recent developments in deep learning i.e. recurrent layers\cite{liang2015recurrent}, residual layers\cite{he2016deep} and use of dense interconnections within convolutional layers\cite{kolavrik2019optimized}. The Recurrent Residual CNN has shown better performance in object detection and can be mathematically expressed as improved residual networks\cite{alom2018recurrent}. The operations of recurrent convolutional layers(RCL) are performed at dicrete time steps where an output depends on previous time step inputs. Let $x_l$ be a input sample in the $l^{th}$ layer of the CNN block and the sample has a pixel location $(i,j)$ on the $k^{th}$ feature maps of the RCL. So the output of the network for that pixel location is given by $y_{ijk}^l(t)$ is given by,
\begin{align}
y_{ijk}^l(t) &= (w_{k}^f)^T * x_{l}^{f(i,j)}(t) + (w_{k}^r)^T * x_{l}^{r(i,j)}(t-1)\\
F(x_l, w_l) &= max(0, y_{ijk}^l(t))
\end{align}
The $F(x_l, w_l)$ represents the output fraom the $l^{th}$ layer after ReLU operation. This output is then passed through the residual unit and the output from the R2CNN is expressed as $x_{l+1}$ and given by,
\begin{equation}
x_{l+1} = x_l + F(x_l, w_l)
\end{equation}
where $x_l$ represents the input samples to the R2CNN layer and it is added with the output from the recurrent layers. In our network inside the recurrent layers and also between the convolutional layers we have included dense interconnections fo better feature representation across layers.  So the output of the network changes to,
\begin{equation}
y_{ijk}^l(t) = (w_{k}^f)^T * H(x_{l}^{f(i,j)})(t) + (w_{k}^r)^T * H(x_{l}^{r(i,j)})(t-1)
\end{equation}
where $H(.)$ is the composite function derived from the concatenation of feature maps from previous convolutional units. 
\newline
The RCL units are used in the basic UNet framework for segmentation using Keras and Tensorflow framework written in Python. The basic U-Net architecture combines a down-sampling (encoder) path to capture the contextual information followed by a symmetrical up-sampling (decoder) path for accurate localization of the features along with skip connections to directly concatenate feature maps from encoder layers to the decoders. Convolutional layers used kernels of size 3X3 with max pooling operation of 2X2 for detection of multiscale features in the encoder portion and deconvolutional layers of kernel size 3X3 was used in the decoder portion of the network. Activation layers after each convolution operation were set as non-linear rectilinear activation units (i.e. ReLU) and a sigmoid function was used for the final activation function setting the network’s output in the range of 0 and 1. In order to mitigate overfitting of the network due to the small dataset size available for training spatial dropouts\cite{hinton2012improving} were implemented in the main architecture and between the recurrent residual convolutional blocks to force the network to efficiently learn the finer image details without overfitting. Binary Crossentropy is used as the loss function and dice coefficient is used as a measure of accuracy for the network. The network architecture is depicted in Fig. \ref{fig:onecol}.
\begin{table*}[]
	
	\label{tab:my-table}
	\centering
	\begin{adjustbox}{width=1.1\textwidth}
		\begin{tabular}{|l|l|l|l|l|l|l|l|l|}
			\hline
			Model        & DSC   & JS & Precision & Recall & Sensitivity & Specificity & Accuracy & AUC \\ \hline
			U-Net        & 0.976 $\pm$ 0.009 &0.956 $\pm$ 0.017    & 0.968 $\pm$ 0.011           & 0.978 $\pm$ 0.017       & 0.990 $\pm$ 0.003             & 0.978 $\pm$ 0.017             &0.990 $\pm$ 0.003          &0.986 $\pm$ 0.008     \\ \hline
			ResUnet      & 0.977 $\pm$ 0.007 &0.957 $\pm$ 0.015   & 0.971 $\pm$ 0.012         & 0.986 $\pm$ 0.014       & 0.991 $\pm$ 0.004           & 0.986 $\pm$ 0.014            &0.990 $\pm$ 0.003          &0.988 $\pm$ 0.006     \\ \hline
			\textbf{Dense R2UNet} & \textbf{0.981 $\pm$ 0.009} &\textbf{0.961 $\pm$ 0.016}    & \textbf{0.982 $\pm$ 0.009}        &  \textbf{0.988 $\pm$ 0.018}      & \textbf{0.994 $\pm$ 0.002}           & \textbf{0.988 $\pm$ 0.018}            & \textbf{0.991 $\pm$ 0.003}         & \textbf{0.989 $\pm$ 0.008}    \\ \hline
		\end{tabular}
	\end{adjustbox}
	\caption[center]{EXPERIMENTAL RESULTS OF THE PROPOSED NETWORK MODEL AGAINST OTHER EXISTING NETWORK MODEL FOR LUNG SEGMENTATION}
\end{table*}
\subsection{Performance Evaluation of the Network}
To assess the segmentation accuracy of the experimental results several performance metrics are considered, like Dice Similarity Coefficient(DSC), Jaccard Score(JS),  Precision, Recall, Sensitivity, Specificity, Accuracy and Area Under the Curve(AUC). To assess these mentioned parameters we calculated the number of True Positive(TP), True Negative(TN), False Positive(FP) and False Negative(FN). The formulas for the parameters are given below,
\begin{align}
DSC &= \frac{2TP}{2TP + FP + FN}\\
JS &= \frac{DSC}{2 - DSC}\\
Precision &= \frac{TP}{TP + FP}\\
Recall &= \frac{TP}{TP + FN} \\
Sensitivity &= \frac{TP}{TP + FN}\\
Specificity &= \frac{TN}{TN + FP}\\
Accuracy &= \frac{TP + TN}{TP + TN + FP + FN}
\end{align}

\begin{figure}[!t]
	\begin{center}
		\includegraphics[width = 23.5em,height=17em]{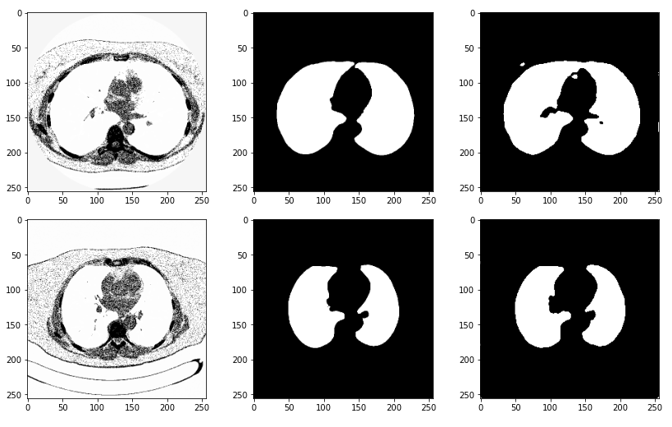} 
	\end{center}
	\caption{Qualitative Assesment of the Dense R2UNet on Lung Segmentation Dataset: 1st column. Input Images, 2nd column. Ground Truth Segmentation and 3rd column. Segmentation output from Dense R2UNet }
	\label{fig:twocol1}
\end{figure}

\section{Experimental Results}
The precise segmentation results achieved with the proposed Dense R2UNet model is shown in Fig. \ref{fig:twocol1} where it is compared to the ground truth. Fig. \ref{fig:twocol} shows the training and validation accuracy of the different models using the Lung Segmentation dataset. These figures show that the proposed Dense R2UNet model has better performance during both training and validation phase when compared to traditional UNet and Residual UNet. Table I shows the summary of how well the Dense R2UNet performs with respect to UNet and ResUNet. It outperforms both other networks in every metrics used to assess segmentation accuracy.

\section{Conclusion and Future Works}
In this paper we proposed a new convolutional neural network architecture based on the traditional UNet model utilizing recurrent residual units and dense interconnections. We refer the new network as Dense R2UNet. The proposed network is evaluated using the Lung Segmentation Dataset. From the results above we can conclude that the Dense R2UNet demonstrates better performance in segmentation tasks with almost same number of network parameters when compared against Traditional UNet and Res-UNet. The dense interconnections introduced direct connections between two layers of same feature map size hence facilitated better propagation of features across the network. The network not only performs better than its counterparts during training but also outbids the others in testing and validation phase. In the future we would like to test the network performance in various other benchmarking datasets to analyze its performance.

\section*{Acknowledgments}
The author wants to thank Dr. Ayan Chakrabarti for his guidance and insightful discussions during the compeltion of the project. I also want to thank PhD mentor Dr. Kooresh Issac Shoghi from the Department of Radiology for providing the GPU resources required to finish this project. I also extend my thanks to my lab members Dr Sudipta Ray and Dr Timothy Whitehead who sacrificed their GPU time to let me use the resource.
{\small
\bibliographystyle{ieee}
\bibliography{refs} 
}

\end{document}